# Measurement and assignment of E-symmetry states in the 6010-6110 cm$^{-1}$ and 8940-9150 cm$^{-1}$ ranges of methane using optical frequency comb double-resonance spectroscopy

Adrian Hjältén[1], Vinicius Silva de Oliveira[1], Michael Rey[2,a], Isak Silander[1], Kevin K. Lehmann[3], and Aleksandra Foltynowicz[1,*]

[1] Department of Physics, Umeå University, 901 87 Umeå, Sweden

[2] Laboratoire Interdisciplinaire Carnot de Bourgogne, UMR CNRS 6303, Université Bourgogne Europe, 9 Av. A. Savary, BP 47870, 21078 Dijon Cedex, France

[3] Departments of Chemistry & Physics, University of Virginia, Charlottesville, VA 22904, USA

*aleksandra.foltynowicz@umu.se

**Abstract**

We use sub-Doppler optical-optical double-resonance (OODR) spectroscopy with a 3.3 μm single-frequency pump and a cavity-enhanced 1.65 μm comb probe to measure 33 ladder-type ($3\nu_3 \leftarrow \nu_3$) and 8 V-type ($2\nu_3$) transitions in the 5880-6090 cm$^{-1}$ range of methane, reaching states with rotational E symmetry in the region of the $P$6 and $P$4 polyads, respectively. We assign the ladder-type transitions using new Hamiltonian predictions and the ExoMol line list, and the V-type transitions using the new Hamiltonian, ExoMol, HITRAN2020, and the WKLMC line lists. While 7 of the states in the $3\nu_3$ range have been previously observed either in earlier OODR work (without cavity enhancement) with 1.5 MHz accuracy or in FTIR measurements of cold bands with 150 MHz resolution, the states reported here have uncertainties down to 150 kHz (5 × 10$^{-6}$ cm$^{-1}$). The E-symmetry states exhibit first-order Stark splitting, which will be reported in our future work.

## 1. Introduction

Methane is a molecule of fundamental interest because it is the simplest molecule with tetrahedral symmetry. It is also a potent greenhouse gas, and it has been detected in the atmospheres of planets in our solar system [1] and on exo-planets [2]. Because of the near coincidence of the normal mode frequencies ($\nu_1 \sim \nu_3 \sim 2\nu_2 \sim 2\nu_4$) and strong couplings between them, the energy levels of methane form groups of strongly interacting states known as polyads. The $P$N polyad is a set of states for which $2n_1 + n_2 + 2n_3 + n_4 = N$, where $n_i$ is the number of quanta in vibrational normal mode i. The complexity of spectra increases with the polyad number, and information is scarce about highly excited states in $P$6 and above. While high resolution laser-based studies exist up to the icosad ($P$5) region [3-8], above 8000 cm$^{-1}$ most of the available line lists are from low- and room-temperature FTIR spectroscopy [9-12].

Optical-optical double-resonance spectroscopy is a perfect tool for selective measurement and assignment of hot-band molecular transitions [8, 13-16]. When the pump frequency is lower than the probe's, two types of OODR transitions are observed: the ladder-type that start from the upper state of the pump transition and reach highly excited levels, and V-type that share the lower state with the pump transition. If the pump is monochromatic, the OODR transitions are free of Doppler broadening, which allows distinguishing them from Doppler-broadened transitions from thermally populated levels in the ground state and from excited states populated by elastic collisions (known as four-level double resonance transitions). We recently developed an OODR spectrometer with a 3.3 μm continuous-wave pump and a comb probe centered around 1.65 μm and used it to record accurate sub-Doppler spectra of methane transitions over a broad bandwidth in the range of the $3\nu_3 \leftarrow \nu_3$ and $2\nu_3$

---

[a] Previously at Groupe de Spectrométrie Moléculaire et Atmosphérique, UMR CNRS 7331, BP 1039, F-51687 Reims Cedex 2, France

bands. In the first demonstration, the sample was contained in a liquid nitrogen-cooled single-pass cell and the frequency accuracy was limited to 1.5 MHz by the frequency drift of the Lamb-dip-locked pump laser [17, 18]. Later, we introduced an enhancement cavity for the comb probe, which reduced the influence of the pump drift on the probe transition frequencies (since the probe is both co- and counter-propagating with the pump) [19]. Most recently, we expanded the spectral coverage of the probe and introduced comb referencing for the pump [20]. So far, we detected 203 ladder-type and 48 V-type transitions, most of which reached states with rotational symmetry $A_1$ [20], $A_2$ [17, 21], $F_1$ [17], $F_2$ [17, 21]; only 5 states (3 reached by a ladder-type transition, and 2 by a V-type) had E symmetry, and they all come from the early version of the spectrometer with liquid-nitrogen cooled single-pass cell and 1.5 MHz accuracy [17].

Here we use the most recent implementation of the cavity-enhanced comb-based OODR spectrometer [20] to measure states with rotational E symmetry in the range of the $P6$ and $P4$ polyads ($3\nu_3$ and $2\nu_3$ bands). This is motivated by the planned experiment to measure the dipole moments of states in these polyads of methane via the Stark splitting. The Stark effect in excited vibrational states of $CH_4$ is dominated by vibrational averaged dipole moments of triply degenerate vibrational states, and thus measurement of such moments should provide a sensitive test of the vibrational character, often highly mixed, predicted by high level calculation of the rovibrational states in this energy region. Due to the small value of these dipole moments (~0.01 in atomic units), experimental considerations suggest that we focus on states with first-order Stark splitting, which in the case of $CH_4$ are rovibrational states of E symmetry. Given the E ↔ E selection rule for dipole transitions in $T_d$ molecules, to reach states of E symmetry one must start in a thermally populated state of E symmetry. The lowest such states are $J$ = (2, E) and (4, E). Final states of lower $J$ value have larger Stark splittings and lower number of $M$ components in the spectrum, so we focused the measurement on OODR spectra starting in the $J$ = (2, E) state.

We pumped the Q(2,E) [$J$=2E(7)← $J$=2E(1)] and R(2,E) [$J$=3E(8)← $J$=2E(1)] transitions of the $\nu_3$ fundamental band with transition frequencies 3018.591351824(77) cm$^{-1}$ [3] and 3048.169096808(97) cm$^{-1}$ [22], respectively. We did not pump the P(2,E) transition because the P(1,E) ladder-type probe transitions reach final states with $J$ = 0, which do not have first order Stark effect, while the upper states of the Q(1,E) and R(1,E) probe transitions should appear in the Q(2,E)-pumped spectrum as the upper states of the P(2,E) and Q(2,E) probe transitions, respectively. Moreover, we already measured the strongest transitions in the P(2,E)-pumped spectrum in our previous work [17].

We assign the measured hot-band (ladder-type) transitions using the new effective Hamiltonian predictions [23] and the ExoMol linelist [24]. For the V-type transitions, we also use the HITRAN2020 [25] and WKLMC line lists [26].

## 2. Experimental setup and procedures

The experimental setup is described in detail in Ref. [20] and is only briefly summarized here. It consists of a high-power 3.3 μm continuous-wave pump produced by an optical parametric oscillator (TOPTICA, TOPO) and a 250 MHz optical frequency comb probe (Menlo Systems FC-1500-250-WG) frequency shifted to around 1.7 μm. Both beams are coupled colinearly into a 60-cm long Fabry-Perot enhancement cavity, which is resonant for the probe with a finesse of ~1100. The mirror reflectivity for the pump is only 2.6%, so it can be treated as single pass. The pump and probe beams are combined in front of the cavity and separated behind the cavity using dichroic mirrors, and the pump beam can be blocked using a shutter placed before the combining dichroic mirror. The comb is locked to the cavity via feedback to the repetition rate, $f_{rep}$, using the Pound-Drever-Hall (PDH) locking method [27], while the carrier-envelope-offset $f_{ceo}$ is stabilized to an RF frequency and adjusted for maximizing the spectral bandwidth of the cavity transmission. Absolute stabilization of $f_{rep}$ is achieved by stabilizing the cavity length using a piezoelectric transducer (PZT). In the measurements presented here, the spectral coverage of the probe was 5880-6080 cm$^{-1}$ (at -10 dB) and the PDH locking point was around 6037 cm$^{-1}$. In the Q(2,E)-pumped measurement, we locked the pump frequency to a Lamb dip of the transition using frequency-modulation (FM) spectroscopy in a 30-cm long reference cell, while for the R(2,E)-pumped measurement we locked the pump to the transition frequency obtained from Ref. [22] using a mid-infrared frequency comb as reference.

The sample pressure for the Q(2,E)-pumped spectrum was 191 mTorr. For the R(2,E)-pumped spectrum we reduced the pressure to 10.3 mTorr to enable the detection of the $2\nu_3$ R(2,E) V-type transition, which would be saturated (i.e. it would absorb all comb light) at the higher pressure. The position of this line can be compared to the accurate measurement from Votava *et al*. [7]. The pump power incident on the sample and the pump transmission on resonance were 880 mW and 31%, and 760 mW and >99% for the Q(2,E) pump and R(2,E) pump, respectively.

The probe transmission through the cavity is analyzed using a Fourier transform spectrometer (FTS). The FTS has a nominal resolution matched to $f_{rep}$ and the two outputs are detected in an auto-balanced configuration by a pair of InGaAs detectors. The optical-path-difference (OPD) is calibrated using a frequency-stabilized HeNe laser (Sios, SL/02/1) with a wavelength $\lambda_{ref} \approx 632.991$ nm. We recorded sequences of interferograms at 130 different $f_{rep}$ values separated by 2.75 Hz, translating to 2 MHz steps of the comb modes in the optical domain. For each $f_{rep}$ step, we acquired one spectrum with pump excitation and one background spectrum with the pump blocked using the shutter. The $f_{rep}$ scans were repeated in alternating directions for averaging of the spectra. The total number of $f_{rep}$ scans was 9 when pumping the Q(2,E) transition and 10 when pumping the R(2,E) transition. We set the relative polarization of the pump and probe lasers to 54.7°, the so called magic angle, which allows measuring the intrinsic probe line intensities without effects due to pump-induced $M_j$ alignment [28].

## 3. Spectral analysis

We use the sub-nominal sampling-interleaving technique [29, 30] to achieve comb-mode limited resolution in the FTS spectra. The spectral analysis follows that described in Refs. [20, 21] and is also only briefly summarized here. To detect the OODR transitions we normalized the spectra with the pump excitation to the corresponding background spectra and interleaved them to a point spacing of 2 MHz. The background normalization largely cancels the Doppler-broadened lines as well as the shape of the comb envelope. The remaining baseline was modeled as a 5$^{th}$ order polynomial and a sum of sine terms prior to interleaving. We used a peak finding routine described in Ref. [21] to detect the sub-Doppler OODR V-type and ladder-type features.

### 3.1. Fitting of the OODR transitions

We analyzed the ladder-type lines in the background-normalized spectra using the procedure described in Refs. [20, 21]. We modeled them in the cavity transmission function [31] as a sum of a sub-Doppler Lorentzian component and a broader Gaussian component to account for collision induced velocity redistribution. The model also included the absorption and dispersion of the Doppler-broadened background simulated using HITRAN2020 parameters [25]. We fitted the lines in windows of ±450 MHz. The free parameters were the integrated absorption and the widths of the Lorentzian and Gaussian components, a common center frequency of both features, and the comb-cavity offset, originating from cavity mirror dispersion [31]. The cavity enhancement factor was fixed to values determined from the Doppler broadened lines using the method described in Ref. [20]. For lines where the SNR of the Gaussian component was below 5, we fixed the Gaussian width and the ratio of the integrated absorption of the Gaussian and Lorentzian components to mean values obtained from the remaining stronger lines in each measurement. In the Q(2,E)-pumped spectrum, sidebands caused by the 25 MHz phase modulation of the pump laser used for Lamb-dip locking were visible at ±50 MHz. The factor of two difference from the modulation frequency stems from the wavenumber ratio of the probe and pump. We modeled the sidebands as two additional Lorentzian shapes at frequencies calculated from the wavenumbers of the pump and probe transition and the pump modulation frequency. Their intensities were fixed to a fraction of the integrated absorption of the center Lorentzian peak that was optimized by manual adjustment and visual inspection on the line with the clearest sidebands. The fit also included a baseline, in most cases a 1st order polynomial, though in cases with more pronounced baseline structure, a 3rd order polynomial was used. In some cases, we also reduced the width of the fit window to avoid baseline issues on the wings of the lines. A ladder-type line observed at 6027.00306(1) cm$^{-1}$ in the Q(2,E)-pumped measurement is shown in Figure 1a) (black) together with the fit (red) and the Lorentzian (green) and Gaussian (blue) fit components. The bottom panel shows the fit residuals.

For fitting of the V-type features, we interleaved spectra without background normalization in order to retain the Doppler-broadened lines. Following the same procedure as in Refs. [20, 21] we removed the baseline caused by the comb envelope by modeling the cavity-enhanced Doppler-broadened $CH_4$ absorption of the overtone region using the HITRAN2020 parameters [25]. Prior to fitting a V-type transition, we canceled the surrounding Doppler-broadened lines by division with a model simulated using HITRAN2020 parameters. We then fitted the Doppler-broadened line displaying the V-type feature in a window of ±1 GHz, with the center frequency, integrated absorption and comb-cavity phase as free parameters. Through division by the resulting model, we isolated the V-type absorption dip which we fitted in the same way as the ladder-type transitions, only inverting the sign of both the Lorentzian and Gaussian components. The Gaussian width and the ratio of the Lorentzian and Gaussian integrated absorption were fixed for all lines to the same values as for ladder-type lines with low SNR. The sidebands in the Q(2,E)-pumped spectrum were below the noise level for the V-types and thus omitted from the model. The fit window was nominally ±500 MHz, but was reduced for lines with problematic baseline structure. A 3rd order polynomial baseline was included in all fits. Figure 1b) shows a V-type transition at 5983.18341(1) cm$^{-1}$ (black) after cancelling the Doppler-broadened line, the fit (red) and the individual Lorentzian (green) and Gaussian (blue) fit components. The bottom panel shows the fit residuals.

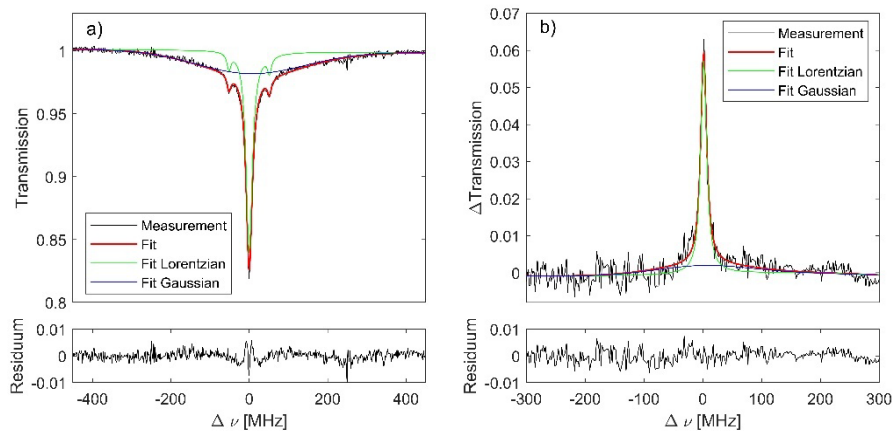

Figure 1. a) A Q(2,E) ladder-type transition (black) at 6027.00306(1) cm$^{-1}$ observed when pumping the $\nu_3$ Q(2,E) transition together with the fit (red) with Lorentzian (green) and Gaussian (blue) component. b) P(2,E) V-type transition (black) at 5983.183412(9) cm$^{-1}$ observed when pumping the $\nu_3$ R(2,E) transition together with the fit (red) with Lorentzian (green) and Gaussian (blue) components. The lower panels display the fit residuals.

## 3.2. Uncertainties

The uncertainty of OODR transition wavenumbers includes contributions from three sources; the optimization of the reference laser wavelength $\lambda_{ref}$ in the FTS, the line fits, and the pressure shift.

The sub-nominal approach [29, 30] requires matching the FTS sampling points to the comb mode frequencies, which is achieved by adjusting the effective reference laser wavelength $\lambda_{ref}$ in post processing to minimize instrumental line shape (ILS) of absorption lines. We determined the optimum $\lambda_{ref}$ as the mean of the optima obtained for a number of ladder-type transitions with highest signal-to-noise ratio (SNR): five lines with SNR > 100 in the Q(2,E)-pumped spectrum, and four lines with SNR > 30 in the R(2,E)-pumped spectrum) with. The uncertainty in $\lambda_{ref}$ was taken as the standard deviation of the optimum values. We estimated the dependence of the retrieved center frequencies on $\lambda_{ref}$ by fitting three strong ladder-type lines in spectra analyzed with a range of $\lambda_{ref}$ around the optimum. We multiplied the largest observed frequency shift per change in $\lambda_{ref}$ by the uncertainty of $\lambda_{ref}$ to obtain the uncertainty contribution to reported center frequencies. This was found to be 290 kHz and 110 kHz for the Q(2,E)-pumped and R(2,E)-pumped measurements, respectively.

The fit uncertainties ranged between 40 kHz and 2.5 MHz for the ladder-type transitions and between 200 kHz and 1.4 MHz for the V-type transitions, depending on the SNR of the lines.

Based on the values reported by Lyulin *et al*. [32] for Doppler-broadened $2\nu_3$ transitions, we calculated the expected self-induced pressure shifts to be 125 kHz at 191 mTorr and 7 kHz at 10.3 mTorr. As discussed in Ref. [20], we consider these values to be overestimated for sub-Doppler transitions, but we conservatively include a contribution equal to these pressure shift estimates in the total uncertainty budget, but do not apply these calculated shifts to the observed transition frequencies.

The uncertainties of the integrated absorption values of the ladder-type lines are a combination of the fit uncertainties and the 5% uncertainty contribution from the determination of the cavity enhancement factor.

## 4. Results

### 4.1. Ladder-type assignment

We retrieved the parameters of 21 ladder-type transitions from the upper state of the $\nu_3$ Q(2,E) transition and 12 from the upper state of the $\nu_3$ R(2,E) transition. Figure 2 depicts the observed ladder-type transitions (black sticks) compared to those predicted by the effective Hamiltonian [23] (red sticks, plotted inverted for clarity) when pumping the a) Q(2,E) and b) R(2,E) transitions. For the experimental data, we plot the integrated absorption, while for the predictions we plot the Einstein *A*-coefficients. The reason why fewer transitions are observed in the R(2,E)-pumped spectrum is the lower sample pressure. Moreover, a few lines are missing because of overlap with the Doppler-broadened overtone lines.

It was straightforward to assign the ladder-type transitions to the Hamiltonian model by comparing the observed and predicted wavenumbers and intensities. In all cases, the observed lines were assigned to the predicted transitions closest in wavenumber. While the agreement with ExoMol was slightly worse, assignment was uncomplicated, particularly having the previous assignments to the Hamiltonian as reference. Figure 3a) and Figure 4a) show comparisons of the experimental transition wavenumbers retrieved when pumping the Q(2,E) (blue) and R(2,E) (red) transitions to the Hamiltonian and ExoMol line lists, respectively. The mean and standard deviation of the discrepancies are given in Table 1. These discrepancies are similar to those observed in our previous work [19, 21].

To be able to compare the experimental line intensities to predictions without knowing the population in the pumped states, we calculated the normalized intensities $I_{N\ obs}$ [17] by taking the ratios of the integrated absorption of the Lorentzian components of the ladder-type transitions and one selected V-type transition for each measurement (indicated in by R and Q in Table 4 of Section 4.2). We calculated the theoretical normalized intensities as $I_{N\ calc} = (A_{LT} g'_{LT}/v_{LT}^2)/(A_{VT} g'_{VT}/v_{VT}^2)$, from the Einstein *A*-coefficients $A_k$, the upper state degeneracies $g'_k$ and the transition wavenumbers $v_k$ of the ladder-type ($k=LT$) and the V-type ($k=VT$) transitions, respectively. Figure 3b) and Figure 4b) show the ratios of observed normalized intensities to those predicted by the Hamiltonian and ExoMol, respectively. The *A*-coefficients for the V-type transitions were taken from HITRAN in both cases. The ratios are scattered around 1 but systematically slightly higher for the Q(2,E) pump line (blue) than for R(2,E) (red). The mean and standard deviation of the intensity ratios are listed in Table 1 individually for each pump transition.

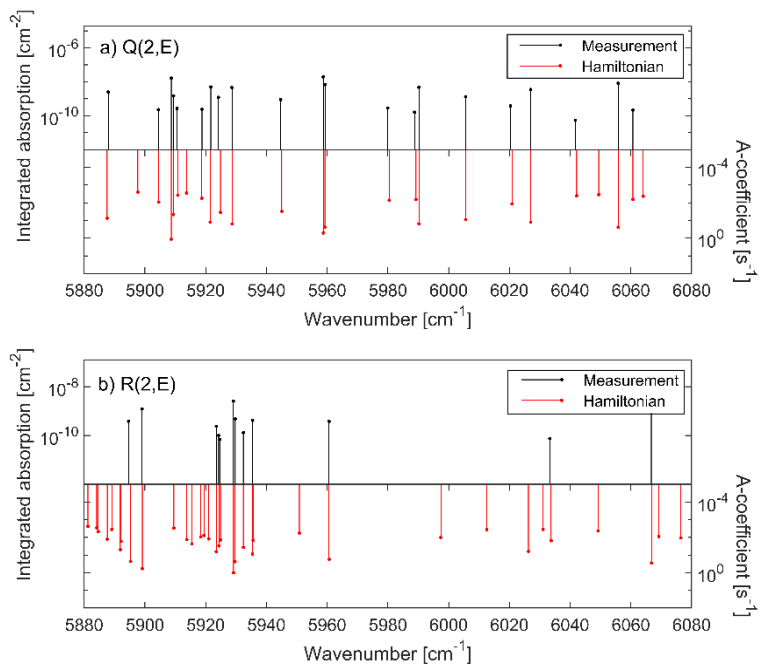

Figure 2. Detected ladder-type transitions (black sticks) and Hamiltonian predictions (red sticks, inverted) represented by their *A*-coefficients in log-scale when pumping the a) Q(2,E) and b) R(2,E) transitions.

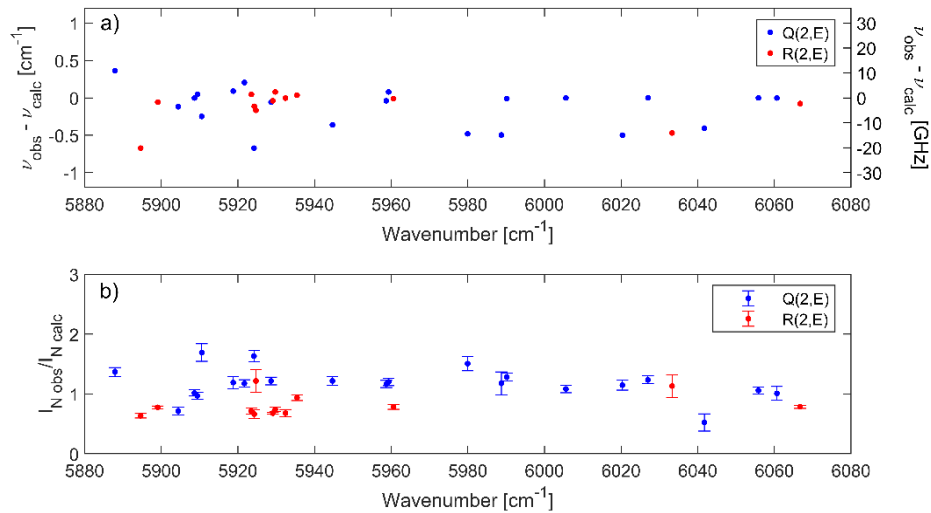

Figure 3. a) The discrepancies in ladder-type wavenumbers between the experiment and Hamiltonian predictions when pumping the Q(2,E) (blue) and R(2,E) (red) transitions. b) The ratios of normalized intensities from the measurements to those calculated from the Hamiltonian predictions when pumping the Q(2,E) (blue) and R(2,E) (red) transitions.

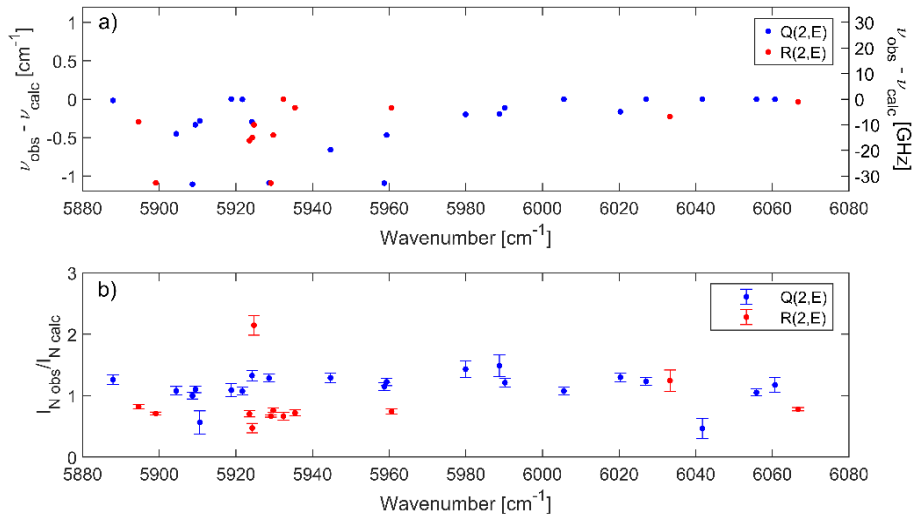

Figure 4. a) The discrepancies in ladder-type wavenumbers between the experiment and ExoMol predictions when pumping the Q(2,E) (blue) and R(2,E) (red) transitions. b) The ratios of normalized intensities from the measurements to those calculated from the ExoMol predictions when pumping the Q(2,E) (blue) and R(2,E) (red) transitions.

Table 1. The mean and standard deviations of the discrepancies in center wavenumber and normalized intensity ratios compared to the Hamiltonian and ExoMol. For normalized intensity ratios the numbers for the Q(2,E) and R(2,E) pump transitions are given separately.

| Reference | $\Delta\nu$ [cm$^{-1}$] | | $I_{N\,obs} / I_{N\,calc}$: Q(2,E) | R(2,E) | |
|---|---|---|---|---|
| | Mean | Std | Mean | Std |
| Hamiltonian | -0.12 | 0.25 | 1.2 | 0.81 | 0.26 | 0.18 |
| ExoMol | -0.34 | 0.37 | 1.1 | 0.87 | 0.24 | 0.44 |

Table 2 presents the parameters and assignments of observed ladder-type transitions sorted by the upper state term value and a comparison to the Hamiltonian predictions. The Supplementary Material contains the assignment from Table 2 together with ExoMol center wavenumbers. We calculated the upper state term values as the sum of our experimental wavenumbers, the pump transition frequencies from Refs. [3, 22] (with kHz accuracy), and the ground state $J = (2, E(1))$ term value of 31.4421218(8) cm$^{-1}$ from Ref. [33] (obtained through private communication with Hiroyuki Sasada). The uncertainties were propagated from all three sources but are dominated by the uncertainties in the line positions from our work. We observed 5 combination differences, i.e. pairs of ladder-type lines that reached the same final state from the two different pumped states. For those lines the reported final state term values are the weighted mean of the values of the two transitions. All upper term values observed twice agreed with their weighted mean within 1σ and the standard deviation of the scatter was 67 kHz. In addition, three of the final states were reached in our previous work [17] via pumping the P(2,E) transitions, and these are indicated by an asterisk in Table 2. Two of these term values agree within 1σ with the values of Ref. [17], which had uncertainties larger by roughly 1 order of magnitude than in the current work, and the third is just outside 1σ. In addition, we found 4 combination differences with the FTIR measurements of cold bands in the 9000 cm$^{-1}$ region by Nikitin *et al.* [11], marked by double asterisk in the Table. The term values of our work were offset from Ref. [11] by on average -40 MHz with a standard deviation of the discrepancies of 26 MHz. This agreement seems plausible considering that the spectral resolution in Ref. [11] was 0.005 cm$^{-1}$ (150 MHz). We note that three E-symmetry states in the *P*6 polyad were reported by Dudás *et al.* [34] from cavity-ringdown spectroscopy of methane in a supersonic flow, but they are outside the range of our spectrometer.

Table 2. Retrieved parameters of the ladder-type transitions (sorted by the upper state term value) and a comparison to the Hamiltonian predictions. Columns: Pump transition, experimental transition wavenumber, observed upper state term value, integrated absorption of the Lorentzian component, Lorentzian width, predicted transition wavenumber, predicted $A$-coefficient, upper state vibrational assignment, upper state $J$-number, upper state counting number. *Levels probed in Ref. [17]. **Levels probed in Ref. [11].

| Pump transition | Experimental transition wavenumber [cm$^{-1}$] | Experimental upper state term value [cm$^{-1}$] | Integrated absorption [10$^{-9}$ cm$^{-2}$] | Lorentzian HWHM [MHz] | Hamiltonian transition wavenumber [cm$^{-1}$] | Hamiltonian Einstein A-coefficient [s$^{-1}$] | Upper vibrational state | Upper state $J$-number | Upper state counting number |
|---|---|---|---|---|---|---|---|---|---|
| Q(2,E) | 5887.94204(1) | 8937.97551(1) | 2.5(1) | 8.8(3) | 5887.577933 | 0.07595 | 0 0 3 0 2F$_2$ | 2E | 251 |
| Q(2,E) | 5904.43799(3) | 8954.47146(3) | 0.23(2) | 7.5(8) | 5904.557534 | 0.009574 | 0 3 1 1 4F$_2$ | 3E | 337 |
| Q(2,E) | 5908.68270(1) | 8958.71617(1) | 16.3(8) | 8.82(8) | 5908.684606 | 1.122 | 0 0 3 0 1F$_1$ | 1E | 150 |
| Q(2,E) | 5909.44048(1) | 8959.47396(1) | 1.47(8) | 7.7(1) | 5909.393127 | 0.04534 | 0 3 1 1 2F$_2$ | 3E | 339 |
| Q(2,E) | 5910.58049(3) | 8960.61396(3) | 0.27(2) | 7.1(7) | 5910.829075 | 0.004783 | 0 3 1 1 1F$_1$ | 3E | 340 |
| Q(2,E) | 5918.80047(2) | 8968.83394(2) | 0.24(2) | 8.4(6) | 5918.710109 | 0.006015 | 1 2 1 0 1F$_2$ | 3E | 344 |
| Q(2,E) | 5921.70189(1) | 8971.73537(1)** | 5.0(3) | 7.96(7) | 5921.496781 | 0.127 | 0 0 3 0 1F$_2$ | 3E | 345 |
| Q(2,E) | 5924.21300(2) |  | 1.21(8) | 9.8(7) | 5924.886318 | 0.0313 |  |  |  |
| R(2,E) | 5894.63524(2) | 8974.24646(1) | 0.38(3) | 6.6(5) | 5895.308258 | 0.2183 | 0 2 2 0 1A$_1$ | 2E | 254 |
| Q(2,E) | 5928.66563(1) |  | 4.5(2) | 8.46(7) | 5928.723594 | 0.1574 |  |  |  |
| R(2,E) | 5899.087883(5) | 8978.699102(5)* | 1.23(6) | 6.1(1) | 5899.145534 | 0.5785 | 0 0 3 0 1F$_1$ | 2E | 255 |
| Q(2,E) | 5944.68698(1) | 8994.72045(1) | 0.89(5) | 8.2(3) | 5945.048784 | 0.03093 | 0 5 0 1 1F$_1$ | 2E | 258 |
| R(2,E) | 5923.53735(1) | 9003.14857(1) | 0.24(1) | 6.3(3) | 5923.491264 | 0.06801 | 0 3 1 1 2F$_2$ | 4E | 441 |
| R(2,E) | 5924.27906(2) | 9003.89027(2) | 0.099(8) | 5.3(5) | 5924.392946 | 0.03058 | 0 3 1 1 4F$_1$ | 4E | 442 |
| R(2,E) | 5924.72070(3) | 9004.33191(3) | 0.069(7) | 7.2(8) | 5924.887082 | 0.01165 | 1 2 1 0 1F$_1$ | 4E | 443 |
| Q(2,E) | 5958.72634(1) |  | 20(1) | 9.21(8) | 5958.766051 | 0.5103 |  |  |  |
| R(2,E) | 5929.148592(4) | 9008.759811(4) | 2.6(1) | 6.05(4) | 5929.187991 | 0.9948 | 0 0 3 0 1F$_1$ | 3E | 349 |
| Q(2,E) | 5959.33938(1) |  | 6.8(3) | 8.42(7) | 5959.26035 | 0.2407 |  |  |  |
| R(2,E) | 5929.761625(5) | 9009.372845(5)* | 0.47(2) | 6.1(1) | 5929.682289 | 0.2345 | 1 4 0 0 1A$_1$ | 2E | 259 |
| R(2,E) | 5932.41759(1) | 9012.02881(1)** | 0.132(8) | 5.9(4) | 5932.421289 | 0.03986 | 1 2 1 0 1F$_2$ | 4E | 447 |
| R(2,E) | 5935.445403(5) | 9015.056622(5) | 0.41(2) | 6.1(1) | 5935.41015 | 0.09081 | 1 2 1 0 1F$_2$ | 4E | 448 |
| Q(2,E) | 5979.95281(2) | 9029.98628(2) | 0.28(2) | 9.4(5) | 5980.433242 | 0.008122 | 1 4 0 0 1E | 2E | 261 |
| Q(2,E) | 5988.76376(8) | 9038.79723(8) | 0.16(4) | 9(2) | 5989.260551 | 0.005904 | 0 2 2 0 1F$_2$ | 2E | 262 |
| Q(2,E) | 5990.20060(1) |  | 4.8(2) | 8.6(1) | 5990.211037 | 0.1617 |  |  |  |
| R(2,E) | 5960.622846(5) | 9040.234066(5) | 0.38(2) | 6.2(1) | 5960.632977 | 0.1807 | 0 2 2 0 1A$_1$ | 2E | 263 |
| Q(2,E) | 6005.59716(1) | 9055.63064(1)* | 1.31(7) | 8.5(2) | 6005.59845 | 0.0875 | 0 0 3 0 2F$_2$ | 1E | 156 |
| Q(2,E) | 6020.34831(2) | 9070.38178(2) | 0.38(2) | 10.1(5) | 6020.847816 | 0.01032 | 0 2 2 0 1F$_2$ | 3E | 359 |
| Q(2,E) | 6027.00306(1) | 9077.03653(1)** | 3.5(2) | 8.58(9) | 6027.00385 | 0.1248 | 0 0 3 0 2F$_2$ | 2E | 266 |
| Q(2,E) | 6041.72234(4) | 9091.75581(4) | 0.05(1) | 3(1) | 6042.127545 | 0.004571 | 0 2 2 0 2A$_1$ | 2E | 269 |
| Q(2,E) | 6055.83452(1) | 9105.86799(1) | 8.2(4) | 8.1(1) | 6055.83545 | 0.2452 | 0 0 3 0 2F$_2$ | 3E | 365 |
| Q(2,E) | 6060.66191(5) | 9110.69538(5)** | 0.22(3) | 8(1) | 6060.66395 | 0.006836 | 0 0 3 0 2F$_2$ | 3E | 366 |
| R(2,E) | 6033.29464(7) | 9112.90586(7) | 0.074(9) | 12(2) | 6033.764667 | 0.01396 | 0 2 2 0 1F$_2$ | 4E | 468 |
| R(2,E) | 6066.718010(8) | 9146.329229(8) | 1.05(6) | 6.6(2) | 6066.795044 | 0.2861 | 0 0 3 0 2F$_2$ | 4E | 475 |

## 4.2. V-type assignment

We retrieved parameters of 8 V-type transitions, 5 in each measurement with two being observed for both pump transitions. For those two, we calculated the weighted mean of their center wavenumbers, and the uncertainty as the standard error of the weighted mean. Two of the transitions were observed previously in Ref. [17] and the wavenumbers agreed within 3 MHz, which is the uncertainty of values reported in Ref. [17]. The $2\nu_3$ transition at 6036.65388(1) cm$^{-1}$ was measured with sub-kHz accuracy by Votava *et al*. [7] and our value agrees with that measurement to within 16 kHz with an uncertainty of 300 kHz. Assignment to the effective Hamiltonian predictions, ExoMol, HITRAN2020 and the experimental WKLMC line list of Nikitin *et al*. [26] was straightforward, knowing the lower state of the transitions. Figure 5 shows a comparison of the transition wavenumbers of the V-type transitions to the Hamiltonian (black crosses), HITRAN (red triangles) ExoMol (green squares) and WKLMC (blue circles) line lists. The mean and standard deviation of the discrepancies are summarized in Table 3. Table 4 gives the retrieved center wavenumbers and assignments of the V-type transitions together with the center wavenumbers and line intensities predicted by the Hamiltonian. The transitions used for calculating normalized ladder-type intensities are indicated in the second column, and those previously measured in Ref. [17] are indicated by an asterisk. Comparison to other line lists is shown in the Supplementary Material.

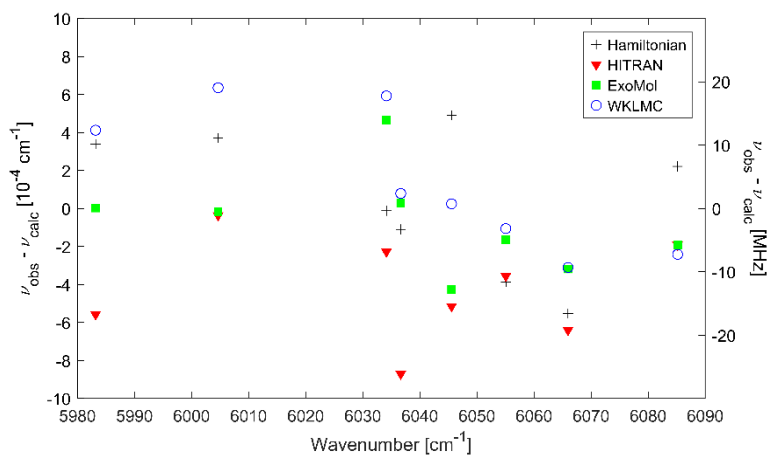

Figure 5. The discrepancies in transition wavenumbers of V-types compared to the Hamiltonian predictions (black crosses), and the HITRAN2020 (red triangles), ExoMol (green squares) and WKLMC (blue circles) line lists.

Table 3. The mean and standard deviation of the discrepancies in V-type line center wavenumbers compared to the Hamiltonian, HITRAN, ExoMol and WKLMC.

| Reference | $\Delta\nu$ [cm$^{-1}$] | |
|---|---|---|
| | Mean | Std |
| Hamiltonian | 4.5·10$^{-5}$ | 3.8·10$^{-4}$ |
| HITRAN | 4.2·10$^{-4}$ | 2.7·10$^{-4}$ |
| ExoMol | -7.8·10$^{-5}$ | 2.7·10$^{-4}$ |
| WKLMC | 1.4·10$^{-4}$ | 3.7·10$^{-4}$ |

Table 4. Retrieved parameters of the V-type transitions and a comparison to the Hamiltonian predictions. Columns: Pump transition(s), observed transition wavenumber, predicted transition wavenumber, predicted line intensity, upper state vibrational assignment, upper state J-number, upper state counting number. $^Q$Used for intensity normalization in Q(2,E)-pumped measurement. $^R$Used for intensity normalization in R(2,E)-pumped measurement. *Levels probed in Ref. [17].

| Pump transitions | Experimental transition wavenumber [cm$^{-1}$] | Hamiltonian transition wavenumber [cm$^{-1}$] | Hamiltonian line intensity [cm/molec] | Upper vibrational state | Upper state J-number | Upper state counting number |
|---|---|---|---|---|---|---|
| R(2,E) | 5983.183412(9)* | 5983.183073 | 1.835e-22 | 0 0 2 0 1F$_2$ | 1E | 30 |
| R(2,E) | 6004.65254(1) | 6004.652165 | 3.241e-22 | 0 0 2 0 1F$_2$ | 2E | 53 |
| Q(2,E), R(2,E) | 6034.14259(4) | 6034.142605 | 1.025e-23 | 0 2 1 0 2F$_2$ | 1E | 31 |
| R(2,E) | 6036.65388(1)$^{R,*}$ | 6036.653989 | 5.041e-22 | 0 0 2 0 1F$_2$ | 3E | 71 |
| Q(2,E) | 6045.55852(2) | 6045.558034 | 9.549e-24 | 0 2 1 0 2F$_2$ | 1E | 33 |
| Q(2,E), R(2,E) | 6055.06999(5)$^Q$ | 6055.070381 | 1.961e-23 | 0 2 1 0 1F$_2$ | 2E | 55 |
| Q(2,E) | 6065.95829(3) | 6065.958843 | 1.741e-23 | 0 2 1 0 2F$_2$ | 2E | 57 |
| Q(2,E) | 6085.16536(3) | 6085.165137 | 2.922e-23 | 0 2 1 0 1F$_2$ | 3E | 73 |

## 5. Conclusions

We report the measurement of 33 ladder-type and 8 V-type transitions in the 5880-6090 cm$^{-1}$ spectral range of methane reaching states with rotational quantum number $J$ =1-4 and E ro-vibrational symmetry in the region of the 3$\nu_3$ and 2$\nu_3$ bands, respectively. Highly-accurate positions of these lines are needed for the measurement of their Stark splitting using a continuous-wave OODR spectrometer described in Ref. [20], which in turn allows extracting the dipole moments for these states. The results of such study are currently being prepared for publication.

The measurements reported here increase the total number of CH$_4$ lines observed using comb-based OODR to 241 and 56 for the ladder- and V-types, respectively, and introduce 25 new upper state term values in the range between 8940 cm$^{-1}$ and 9150 cm$^{-1}$. The sub-MHz accuracy of our measurements is confirmed by comparison with previous measurements of Votava *et al*. [7], and it is better than that of previous single-pass OODR [17] and FTIR measurements [11].

## 6. Acknowledgements

This project is supported by the Knut and Alice Wallenberg Foundation (grant: KAW 2020.0303), and the Swedish Research Council (grant: 2020-00238). K.K.L. acknowledges funding from the U.S. National Science Foundation (grant: CHE-2108458) and the Wenner Gren Foundation (grant: GFOv2024-0010). M.R. acknowledges support from the French National Research Agency TEMMEX project (grant: 21-CE30-0053-01).